\documentclass[twocolumn,prl,reprint,letterpaper]{revtex4-1}

\usepackage{graphicx}
\usepackage{dcolumn}
\usepackage{bm}
\usepackage{amsmath}
\usepackage{amssymb}
\usepackage{amsfonts}
\usepackage{color}
\usepackage{psfrag}

\newcommand{\ud}{\mathrm{d}}

\begin{document}

\preprint{APS/123-QED}

\title{Evolutionary stochastic dynamics of speciation\\ and a simple genotype-phenotype map for protein binding DNA}

\author{Bhavin S. Khatri}
 \email{bhavin.khatri@physics.org}
\author{Richard A. Goldstein}%

\affiliation{%
MRC National Institute for Medical Research\\
Mathematical Biology Division\\
The Ridgeway, London, NW7 1AA, U.K.
}%


%

\date{\today}
\begin{abstract}
Speciation is of fundamental importance to understanding the huge diversity of life on Earth. In contrast to current phenomenological models, we develop a biophysically motivated approach to study speciation involving the co-evolution of protein binding DNA for two geographically isolated populations. Our results predict that, despite neutral diffusion of hybrids in trait space, smaller populations have a higher rate of speciation, due to sequence entropy poising populations more closely to incompatible regions of phenotype space. A key lesson of this work is that non-trivial contributions of sequence entropy give rise to a strong population size dependence on speciation rates.

\end{abstract}

\maketitle


Speciation is thought to underly much of the diversity of life on Earth today. The development of quantitative models that can predict speciation rates will thus allow better understanding of the different factors that maintain bio-diversity along with the processes of extinction and environmental change \cite{CoyneOrrSpeciationBook2004,Rosenzweig2001}. Yet the detailed genetic mechanisms by which distinct species arise is still largely not understood. Darwin \cite{Darwin1859}, despite the title of his magnus opus, struggled to understand how natural selection could give rise to hybrid inviability or infertility. If the hybrid inviability were due to a single locus, how could two species evolve from a common ancestor, since at least one of these species would have to evolve past an inviability bottleneck; for example, if a pair of species, which share common ancestry, are fixed for $aa$ and $AA$ alleles, respectively, and the genotype $aA$ is inviable, how did one of these populations evolve the $Aa$ allele?  It has since been understood that non-linear or epistatic interactions between different loci can give rise to so-called Dobzhansky-M\"{u}ller incompatibilities \cite{Dobzhansky1936a,Muller1942,Bateson1909} between independently evolving lineages. For example, two lines evolving independently through geographic isolation (allopatric evolution) from a common ancestor \textit{ab}, can fix the allelic combinations \textit{aB} and \textit{Ab} respectively, yet the hybrid genotype $AB$ is inviable.


Field data \cite{CoyneOrrSpeciationBook2004,MayrBook1963} suggest that the most dominant form of speciation does indeed involve geographically isolated populations with no or very little gene flow and that the mechanism is commonly via Dobzhansky-M\"{u}ller incompatibilities \cite{Wu1983,Vigneault1986}. There are a number of models of allopatric speciation based on Dobzhansky-M\"{u}ller incompatibilities, which consider independent lineages evolving neutrally or under varying selection pressures on each line \cite{Orr1995,Orr2001,OrrOrr1996,Gavrilets2003}. However, all of these models are essentially phenomenological in their assumptions about the genetic basis of speciation. This is understandable without recourse to very detailed models; however, in this paper we develop a generic approach, which explores universal properties that the genetic basis of traits may have on the process of speciation. In particular, much recent work has shown that in general mappings from genotype to phenotype are non-trivial \cite{Fontana2002,Force1999,Berg2004,Khatri2009}, giving rise to a number of previously unpredicted effects, due to the increasing prominence that sequence entropic effects have for small populations. In this paper, we examine the process of how incompatibilities arise in allopatry, for an abstract, yet biophysically motivated model of transcription factor binding, which accounts for the sequence entropy of binding and ask what population size effects such a genotype-phenotype maps induces. The binding of transcription factors to DNA to control gene expression is arguably one of the most important co-evolving systems for organisms and crucial for their correct development and so makes an ideal first case study to study the affects of genotype-phenotype maps on speciation.

To tackle this question we develop the tools of stochastic dynamics for evolution in the regime where $\mu N\ll 1$, where $\mu$ is the mutation rate and $N$ is the population size, and use it to ask how hybrid fitness and the probability of incompatibilities increases with divergence time between a pair of lines under the same stabilising selection pressure. We find, in contrast to current models, which predict no population size effect \cite{Orr1995,OrrOrr1996,Orr2001, Gavrilets2003}, that despite the two lines diverging neutrally, smaller populations have a higher rate of speciation as the effect of sequence entropy is to bias populations more closely to incompatible regions of phenotype space. We suggest a key lesson of this work, is that even though co-evolving loci on different lineages may diverge in a population-size independent neutral manner, non-trivial contributions of sequence entropy in a stabilising fitness landscape give rise to strong population size dependence.


\section{A Smoluchowski Equation for Evolutionary Stochastic Dynamics}

Natural selection acts on phenotypes, however, in general, there will not be a one-to-one mapping between phenotypes and genotypes, but instead many genotypes can code for the same phenotype. As has been shown, for small population sizes ($\mu N\ll 1$), where populations are largely monomorphic, genotype-phenotype maps can give rise to a bias in evolution towards phenotypes with larger sequence entropy and in equilibrium is described by a balance between a tendency of phenotypes to increase their fitness and at the same time maximise their sequence entropy \cite{Iwasa1988,Sella2005}. This is embodied by the free fitness, $\Phi(\boldsymbol{\xi})=F(\boldsymbol{\xi})+S(\boldsymbol{\xi})/\nu$, where $F$ is the Malthusian fitness \cite{Sella2005} and $S=-\langle \ln p(\boldsymbol{\xi})\rangle_{\boldsymbol{\xi}}$ is the sequence entropy, where $p$ is the equilibrium distribution of $\boldsymbol{\xi}$, the vector of traits and $\nu$ is the Lagrange multiplier or effective inverse temperature of the canonical ensemble, which is linear in population size $N$; for the Wright-Fisher process $\nu=2(N-1)$ and the Moran process $\nu=N-1$ \cite{Sella2005}. In equilibrium,  the free fitness is maximised and the probability distribution of traits is Boltzmann distributed, $p(\boldsymbol{\xi})=\frac{1}{Z}e^{\nu\Phi(\boldsymbol{\xi})}$, where $Z=\int\ud\boldsymbol{\xi}e^{\nu\Phi(\boldsymbol{\xi})}$.

Out of equilibrium, we expect that stochastic dynamics will give rise to: 1) diffusion in phenotype space with diffusion constant $\mu$, where $\mu$ is the mutation rate of all sites contributing to the phenotype; 2) directed motion driven by gradients in the free fitness function with respect to changes in phenotype $\boldsymbol{\xi}$. The flux of probability in phenotype space will then be given by

\begin{equation}\label{Eq:SmoluchowskiFlux}
\boldsymbol{J}=-\tfrac{1}{2}\mu\boldsymbol{\nabla}p(\boldsymbol{\xi}) + \frac{1}{\zeta} p(\boldsymbol{\xi})\boldsymbol{\nabla}\Phi(\boldsymbol{\xi})
\end{equation}
where $\zeta$ is a coefficient representing the strength of evolutionary change in response to an evolutionary force (or gradient in free fitness) and the factor of a $\frac{1}{2}$ for the mutation rate comes from converting from a discrete random walk to a continuous one. In equilibrium, detailed balance requires this flux to be zero, from which it is simple to show that

\begin{equation}\label{Eq:EvolutionaryEinsteinRelation}
\zeta=\frac{2}{\nu\mu},
\end{equation}
which is the evolutionary equivalent of the Einstein relation that relates the friction constant to the diffusion constant of a Brownian particle; here the evolutionary friction constant $\zeta$ is inversely proportional to the mutation rate and population size. We can now express the Smoluchowski Equation in its final form using the continuity equation, $\partial_tp(\boldsymbol{\xi})=-\boldsymbol{\nabla}\cdot\boldsymbol{J(\boldsymbol{\xi})}$\footnote{A more exact treatment using the Kramer's Moyal expansion from a generalised Master Equation gives the same Smoluchowski Equation, but for brevity is not shown.}:


\begin{equation}\label{Eq:SmoluchowskiEqn}
\frac{\partial p}{\partial t}=\tfrac{1}{2}\mu\boldsymbol{\nabla}\cdot\left(\boldsymbol{\nabla}p(\boldsymbol{\xi}) - \nu p(\boldsymbol{\xi})\boldsymbol{\nabla}\Phi(\boldsymbol{\xi})\right).
\end{equation}
This Smoluchowski Equation can be converted to an equivalent set of stochastic differential equations \cite{GardinerBook,vanKampen}

\begin{equation}\label{Eq:SmoluchoskiSDEs}
\frac{\ud\xi_i}{\ud t}=\tfrac{1}{2}\nu\mu\frac{\partial\Phi(\boldsymbol{\xi})}{\partial\xi_i}+\eta_i(t),
\end{equation}
where $i$ corresponds to the $i^{\mathrm{th}}$ trait of $\boldsymbol{\xi}$ and where $\eta_i$ is a white noise Gaussian process with moments $\langle\eta_i(t)\rangle=0$ \& $\langle\eta_i(t)\eta_j(t')\rangle=\mu\delta_{ij}\delta(t-t').$
Eqn.\ref{Eq:SmoluchoskiSDEs}, is a generalisation of the Ornstein-Uhlenbeck process for phenotypic evolution described in \cite{Bedford2009}, but for an arbitrary free fitness landscape and including the correct population size dependence of the strength of the drift term via the Einstein relation Eqn.\ref{Eq:EvolutionaryEinsteinRelation}.


\section{A simple continuous model of transcription factor binding DNA}

The two-state approximation \cite{Hippel1986,Gerland2002a} for transcription factor (TF) binding assumes that amino acid base pair hydrogen-binding energies are approximately additive and that each ‘‘nonoptimal’’ interaction increases the energy of binding by the same amount. This suggests replacing DNA and amino acid sequences by binary strings. The binding energy is simply proportional to the Hamming distance $r=(\boldsymbol{g_1}-\boldsymbol{g_2})\cdot(\boldsymbol{g_1}-\boldsymbol{g_2})$ between a pair of binary sequences, $\boldsymbol{g_1}$, $\boldsymbol{g_2}$. We can assign a fitness $F_r$ to each value of $r$, either arbitrarily or using some knowledge of the biophysics and function of a particular system. On the other hand, the entropy is given directly by the nature of the binary model; there are many sequences that give the same Hamming distance, precisely $\Omega_r=\binom{\ell}{r}\approx2^\ell\sqrt{2/\pi \ell}\exp(-\frac{2}{\ell}(r-\ell/2)^2)$, when $\ell$, the sequence length, is large. So to a good approximation the sequence entropy is quadratic in Hamming distance $r$:

\begin{equation}\label{Eq:MutationalEntropy}
S_r=-\frac{2}{\ell}(r-\ell/2)^2 + \mathsf{const}.
\end{equation}
We see that entropy is maximised for $r=\ell/2$, which corresponds to the Hamming distance that has the largest number of genotypes coding it. In order to take advantage of the effective stochastic dynamics described in the previous section, we replace each sequence $\boldsymbol{g}_i$ with a continuous variable $x_i$ and define a Hamming distance like or binding energy variable as $\xi=|x_1-x_2|$ \footnote{Note that for larger populations discrete dynamics will come into play as on each lineage the populations distributions become confined to within a single hamming distance of the zero hamming distance state. This state has zero degeneracy and so the two lineages can only diverge via substitutions to larger, lower fitness, hamming distances at an exponentially reduced rate compared to neutrality.}. Further, we assume a simple quadratic fitness landscape $F(\xi)=-\frac{1}{2}\kappa_F\xi^2$. With sequence entropy given by Eqn. \ref{Eq:MutationalEntropy}, we have (to within a constant) the free fitness given by $\Phi(\xi)=-\frac{1}{2}\kappa_F\xi^2 -\frac{2}{\ell\nu}(\xi-\ell/2)^2$.


Using Eqn. \ref{Eq:SmoluchoskiSDEs}, treating $x_1$ and $x_2$ as independent variables, we can write down a pair of stochastic differential equations describing the dynamics of the sequence-like variables in a quadratic landscape:

\begin{eqnarray}\label{Eq:SmoluchoskiSDEsQuadLandscape}
\frac{\ud x_1}{\ud t}&=&-\tfrac{1}{2}\nu\mu\kappa(x_1-x_2)+\mu\ \mathrm{sgn}(x_1-x_2)+\eta_1(t),\nonumber\\
\frac{\ud x_2}{\ud t}&=&-\tfrac{1}{2}\nu\mu\kappa(x_2-x_1)+\mu\ \mathrm{sgn}(x_2-x_1)+\eta_2(t),
\end{eqnarray}
where $\kappa=\kappa_F +\frac{4}{\ell\nu}$, which is the curvature in the free fitness landscape and is a sum of the curvatures due to fitness and sequence entropy. In addition, we have used the fact that $\partial_z(|z|-c)^2=2(z-c\ \mathrm{sgn}(z))$, where $\mathrm{sgn}$ is the sign function that returns the sign of the argument.

\section{Analytical calculation of hybrid fitness and the probability of an incompatibility}

From Eqn.\ref{Eq:SmoluchoskiSDEsQuadLandscape}, fitness and sequence entropy balance (maximum in free fitness) for $x_1-x_2=\pm\xi_\infty=\pm\frac{2}{\kappa_F\nu}$, so the free fitness is doubled peaked with a cusp valley at $x_1-x_2=0$, except in the limit of an infinite population size ($\nu\rightarrow\infty$). For the sake of analytical tractability, we can assume that each lineage is always confined to one or the other maximum, such that without loss of generality the initial condition $x_1(0)>x_2(0)$ remains true. This implies we can make the simplification $\xi=x_1-x_2$, from which it follows:

\begin{eqnarray}\label{Eq:SmoluchoskiSDEsQuadLandscapeSimple}
\frac{\ud x_1}{\ud t}&=&-\lambda(x_1-x_2)+\mu+\eta_1(t), \nonumber\\
\frac{\ud x_2}{\ud t}&=&-\lambda(x_2-x_1)-\mu+\eta_2(t),
\end{eqnarray}
where the characteristic relaxation rate of the system is given by $\lambda=\nu\mu\kappa/2$. Note that this system of equations is analogous to the dynamics of a pair of overdamped beads connected by spring with a temperature-dependent elastic constant. To solve these equations, it is straightforward to take the Laplace transform of these equations, solve the resulting matrix equation to give solutions in Laplace space and find the inverse Laplace transform to give

\begin{equation}\label{Eq:x(t)}
\boldsymbol{x}(t)=\mathsf{J}\boldsymbol{x}(0)+\frac{1}{\kappa\nu}(1-e^{-2\lambda t})
\begin{pmatrix}
1 \\
-1 \\
\end{pmatrix}
+\int_0^t\mathsf{J}(t-t')\boldsymbol{\eta}(t')\ud t'
\end{equation}
where $\boldsymbol{x}=(x_1,x_2)^T$, $\boldsymbol{\eta}=(\eta_1,\eta_2)^T$, the matrix $\mathsf{J}$ is given by

\begin{equation}\label{Eq:j(t)}
\mathsf{J}=\frac{1}{2}
\begin{pmatrix}
1+e^{-2\lambda t} & 1-e^{-2\lambda t} \\
1-e^{-2\lambda t} & 1+e^{-2\lambda t}\\
\end{pmatrix},
\end{equation}
and it is understood that the integral of the vector above is an element by element operation. There will be an equivalent set of stochastic differential equations for the second lineage of the same form as Eqn.\ref{Eq:SmoluchoskiSDEsQuadLandscapeSimple} with solution $\boldsymbol{x'}(t)$ given by Eqn.\ref{Eq:x(t)} and binding energy $\xi'=|x_1'-x_2'|$ with an identically distributed noise vector, but uncorrelated with the first lineage. If we let $w=x_1-x_2'$ and $w'=x_1'-x_2$ and the hybrid binding energies be $h=|w|$ and $h'=|w'|$, then it is straightforward to show using Eqns.\ref{Eq:x(t)}, \ref{Eq:j(t)} and the moments of the Gaussian process that $\langle w(t)\rangle=\langle w'(t)\rangle=\langle \xi(t)\rangle=\langle \xi'(t)\rangle$, where

\begin{equation}\label{Eq:<h>(t)}
\langle w(t)\rangle=\xi_0e^{-2\lambda t} +\frac{2}{\kappa\nu}(1-e^{-2\lambda t}).
\end{equation}
We see that for long times ($\lambda t\gg1$), the average binding energy $\langle \xi(\infty)\rangle$ decays from the initial condition to its equilibrium value $\xi_\infty=\frac{2}{\kappa\nu}$ on the timescale $1/(2\lambda)$; in reality, using the full dynamics in Eqn.\ref{Eq:SmoluchoskiSDEsQuadLandscape} this variable would have zero mean in the long time limit, but as we see below the results are insensitive to this error. If we define the vector $\boldsymbol{w}=(w,w')^T$, we find that the covariance matrix $\boldsymbol{\Sigma}=\langle(\boldsymbol{w}-\langle\boldsymbol{w}\rangle)^T(\boldsymbol{w}-\langle\boldsymbol{w}\rangle)\rangle$ is symmetric and has elements


\begin{eqnarray}\label{Eq:<h^2>(t)}
\Sigma_{11}&=& \mu t + \frac{1}{2\kappa\nu}(1-e^{-4\lambda t}),\nonumber\\
\Sigma_{12}&=&= -\mu t + \frac{1}{2\kappa\nu}(1-e^{-4\lambda t}).
\end{eqnarray}
We see for short times ($\lambda t\ll1$), the off-diagonal terms are zero, which suggests the binding energies of the two hybrids are uncorrelated. However, for long times ($\lambda t\gg 1$) they become anti-correlated, which shows, unlike at short times, the probability of both hybrids being incompatible are not independent. Average hybrid fitness is then given by $\langle F_h(t)\rangle=-\frac{1}{2}\kappa_F\langle h^2\rangle=-\frac{1}{2}\kappa_F\langle w^2\rangle$ and we see that on long times it decreases like $\sim\mu t$.

A similar calculation of the variance on each lineage gives,
\begin{equation}\label{Eq:<r^2>(t)}
\langle \xi^2\rangle-\langle\xi\rangle^2=\frac{1}{\kappa\nu}(1-e^{-4\lambda t}),
\end{equation}
which shows that, as expected, the co-evolutionary constraint of the free fitness landscape bounds the variance of binding energies on each lineage. So the variance in the hybrid binding energies is due to a pure diffusive term $\mu t$, which represents how the two lineages diffuse apart by independent mutations, plus a term which represents the saturating growth of variance of each lineage after divergence.


The dynamics of the probability of a DMI for each hybrid, irrespective of whether the other hybrid has a DMI or not, is simply,

\begin{equation}\label{Eq:<I>(t)_defn}
P_I(t)=1-\int_{-\xi^*}^{\xi^*}p(w,t)\ud w,
\end{equation}
where $\xi^*=\sqrt{2|F^*|/\kappa_F}$, where $F^*$ is the threshold fitness below which an incompatibility arises in a hybrid. The variable $w$ is given by the sum of a number of Gaussian processes, so $p(w,t)$ itself must be Gaussian (with $p(h,t)=p(w,t)+p(-w,t)$),
which is completely specified by its mean (Eqn.\ref{Eq:<h>(t)}) and variance $\Sigma_{11}=\langle w^2\rangle-\langle w\rangle^2$, which can be directly calculated from Eqns.\ref{Eq:<h>(t)} \& \ref{Eq:<h^2>(t)}. From Eqn \ref{Eq:<I>(t)}, the probability of a DMI is then simply an integral of a Gaussian, which is expressed in terms of complementary error functions:


\begin{eqnarray}\label{Eq:<I>(t)}
P_I(t)&=&\frac{1}{2}\mathrm{erfc} \left(\frac{\xi^* -\langle w\rangle}{\sqrt{2\Sigma_{11}}}\right) +\frac{1}{2}\mathrm{erfc}\left(\frac{\xi^* +\langle w\rangle}{\sqrt{2\Sigma_{11}}}\right).
\end{eqnarray}
Note that both the average hybrid fitness and the probability of incompatibilities (Eqn.\ref{Eq:<I>(t)}) collapse to the same curves as a function of $\mu t$, for the same value of $\kappa\nu=\kappa_F\nu+4/\ell$ and when fitness is measured relative to the curvature $F/\kappa_F$.

\section{Results}

We can compare the analytical calculations of the previous section with numerical simulations of Eqn. \ref{Eq:SmoluchoskiSDEsQuadLandscape} for a pair of lines diverging from a common ancestor, where no approximation is made regarding the values of $x_1$ and $x_2$. The results of the simulations are averaged over $10^4$ independent realisations and all results assume an effective sequence length $\ell=10$, $F^*/\kappa_F=-25$ and $\xi(0)=\xi_\infty=\frac{2}{\kappa\nu}$, which is the mean value of $\xi$ in equilibrium.

\begin{figure}[h!]
\begin{center}
{\rotatebox{0}{{\includegraphics[width=0.5\textwidth]{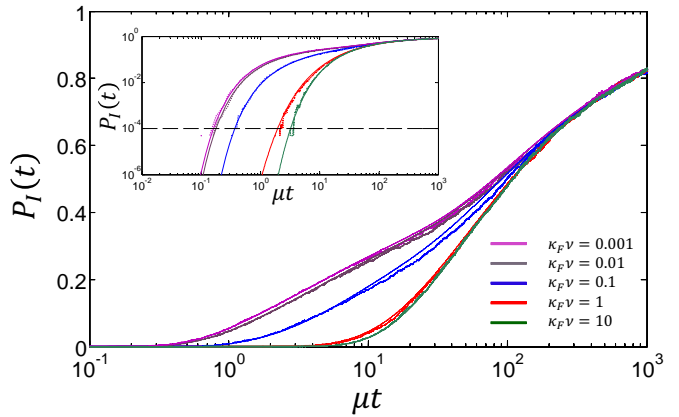}}}}
\caption{Plot of the probability of a DMI $P_I(t)$ for a single hybrid as a function of time for dimensionless population sizes $\kappa_F\nu=\{0.001,0.01,0.1,1,10\}$. Solid lines are the approximate analytical calculations using Eqn.\ref{Eq:<I>(t)} and dotted lines are numerical integration of Eqn.\ref{Eq:SmoluchoskiSDEsQuadLandscape} using Eqn.\ref{Eq:<I>(t)_defn}. The inset shows the same plot on a log-log scale with the dashed line indicating the threshold probability for speciation assuming $M=10000$ equivalent clusters of loci.
 \label{Fig:<I(t)>}}
\end{center}
\end{figure}

In Fig.\ref{Fig:<I(t)>}, we have plotted the probability of an incompatibility, for various values of $\kappa_F\nu$, where solid lines are the analytical calculation and the dotted lines are from the numerical integration of Eqn.\ref{Eq:SmoluchoskiSDEsQuadLandscape}. Firstly, we see that the analytical predictions compare very well to integrating the full stochastic differential equations, validating our simplifying assumptions. Secondly, we see there is a large population size effect for the probability of an incompatibility; the characteristic time for incompatibilities to arise is shorter as the population size decreases, where in addition for small populations ($\kappa_F\nu\ll 0.1$) there are two such characteristic timescales. In addition, we see that the dynamics of $P_I(t)$ become insensitive to changes in population size both for small population sizes ($\kappa_F\nu\ll 0.1$) and large population sizes ($\kappa_F\nu\gg 0.1$). Note that, as we have not bounded the binding energies to a maximum value of $\ell$, we would expect real sequences to have a smaller plateau value of $P_I$ in the long time limit of $\mu t\gg1$; as we argue below it is the short time limit that is most relevant to speciation.
%
%

\begin{figure}[h!]
\begin{center}
{\rotatebox{0}{{\includegraphics[width=0.5\textwidth]{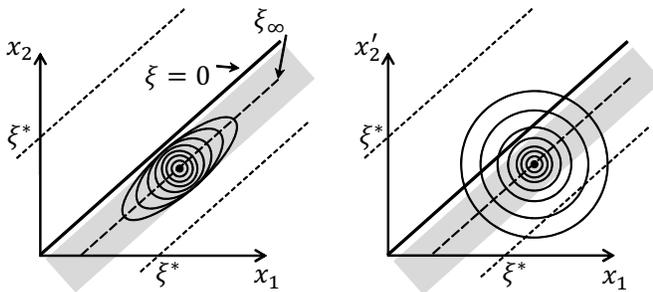}}}}
\caption{Diagram showing the evolution of the probability density of $p(x_1,x_2,t)$ (left) and $p(x_1,x_2',t)$ (right), as a function of time.
 \label{Fig:DMI_Diagram}}
\end{center}
\end{figure}

To understand this general behaviour, we can consider what happens in the phase space of $x_1$ and $x_2$ for the first lineage, versus the phase-space for $x_1$ and $x_2'$ for the hybrids, as shown in Fig.\ref{Fig:DMI_Diagram}. Initially, both lineage and hybrid populations diffuse in a neutral and spherically symmetric manner (like $2\mu t$ and $4\mu t$, respectively) up to the time $\sim\lambda^{-1}$, when the change in free fitness is of order the mean fitness $\sim 1/\nu$ and the accumulated variance of the hybrid population approaches the characteristic width of the potential $\delta\xi^2\sim\frac{1}{\kappa\nu}$. After this time the co-evolutionary constraint of the free fitness landscape is felt on each lineage and the probability density is then squeezed along a tube whose axis is defined by $x_1=x_2+\xi_\infty$ (assuming an initial condition $x_1(0)>x_2(0)$) and width $\delta\xi$. As the marginal pdfs for $x_1$ and $x'_2$ will be identical, the result is that in the hybrid phase-space, we have that the $p(x_1,x'_2,t)$ still grows in a spherically symmetric manner, as indicated by the term $2\mu t$ in Eqn.\ref{Eq:<h^2>(t)}; incompatibilities arise when hybrid populations have diffused to one or the other critical binding energy at $x_1-x_2'=\pm\xi^*$. From Eqn.\ref{Eq:<I>(t)} and Fig.\ref{Fig:DMI_Diagram}, we see that there will in general be two characteristic times for DMIs to arise, given by the condition that $(\xi^*\pm\frac{2}{\kappa\nu})^2\sim \Sigma_{11}(t)$. It is then simple to see that in the limit of $\kappa_F\nu\gg4/\ell$, $\xi_0\rightarrow 0$ and the time to diffuse to each boundary is the same as observed in Fig.\ref{Fig:<I(t)>}.  Finally, for small population sizes ($\kappa_F\nu\ll4/\ell$) $\kappa\nu\rightarrow 4/\ell$, which shows the dynamics of DMIs becomes independent of population size in this limit as well. It is clear that the population size dependence described here is likely to be also seen in more complex models of co-evolving loci as for a general free fitness landscape, the balance between sequence entropy and fitness is population size dependent, poising populations nearer or further away from such incompatible regions.

For the actual process of speciation, it will typically be the short-time behaviour that will dominate. Given a large number $M\sim 1000\rightarrow 10000$ of similar non-interacting clusters of loci, we would expect speciation when $MP_I\gg1$ and so the critical probability of a DMI will be $\sim1/M\sim10^{-4}\rightarrow 10^{-3}$. As shown in the inset of Fig.\ref{Fig:<I(t)>}, which is a plot of $P_I(t)$ on a log-log scale together with the threshold probability $1/M=10^{-4}$, there is strong population size dependence on the time to speciation. In particular, we see this time is relatively insensitive to the exact value of $M$, compared to the population size effect we describe and arises due to the steep gradient of $P_I(t)$ at short times.

The process of speciation underlies the vast diversity of life on Earth. However, conventional models of speciation via Dobzhansky-M\"{u}ller incompatibilities for geographically isolated populations with neutral divergence predict that speciation is independent of population size \cite{Orr1995,OrrOrr1996,Orr2001, Gavrilets2003}. However, such models are phenomenological with respect to the genetic basis of incompatibilities. Here using the tools of stochastic dynamics and a simple biophysically motivated model for the genotype-phenotype map of protein binding DNA, we have shown that there is a significant population size dependence of the time for hybrid incompatibilities to arise for independently evolving isolated populations under the same stabilising selection pressure. Protein binding DNA to control gene expression is a prototypical co-evolving system and critical for the proper development of organisms, thus these results have strong implications for the speciation rates and diversity of populations at small population sizes. We should note that our model does not account for differential selection across lineages, although it can easily be extended to do so, which can reinforce or have an opposite dependence on population size dependent on whether selection is beneficial or mildly deleterious \cite{Gavrilets2003}; the work here highlights a previously unanticipated population size dependence, which is necessary for a fully quantitative model of speciation.



\begin{acknowledgements}
We acknowledge useful discussions with David Pollock, University of Colorado and funding from the Medical Research Council, U.K.
\end{acknowledgements}


\begin{thebibliography}{26}%
\makeatletter
\providecommand \@ifxundefined [1]{%
 \@ifx{#1\undefined}
}%
\providecommand \@ifnum [1]{%
 \ifnum #1\expandafter \@firstoftwo
 \else \expandafter \@secondoftwo
 \fi
}%
\providecommand \@ifx [1]{%
 \ifx #1\expandafter \@firstoftwo
 \else \expandafter \@secondoftwo
 \fi
}%
\providecommand \natexlab [1]{#1}%
\providecommand \enquote  [1]{``#1''}%
\providecommand \bibnamefont  [1]{#1}%
\providecommand \bibfnamefont [1]{#1}%
\providecommand \citenamefont [1]{#1}%
\providecommand \href@noop [0]{\@secondoftwo}%
\providecommand \href [0]{\begingroup \@sanitize@url \@href}%
\providecommand \@href[1]{\@@startlink{#1}\@@href}%
\providecommand \@@href[1]{\endgroup#1\@@endlink}%
\providecommand \@sanitize@url [0]{\catcode `\\12\catcode `\$12\catcode
  `\&12\catcode `\#12\catcode `\^12\catcode `\_12\catcode `\%12\relax}%
\providecommand \@@startlink[1]{}%
\providecommand \@@endlink[0]{}%
\providecommand \url  [0]{\begingroup\@sanitize@url \@url }%
\providecommand \@url [1]{\endgroup\@href {#1}{\urlprefix }}%
\providecommand \urlprefix  [0]{URL }%
\providecommand \Eprint [0]{\href }%
\providecommand \doibase [0]{http://dx.doi.org/}%
\providecommand \selectlanguage [0]{\@gobble}%
\providecommand \bibinfo  [0]{\@secondoftwo}%
\providecommand \bibfield  [0]{\@secondoftwo}%
\providecommand \translation [1]{[#1]}%
\providecommand \BibitemOpen [0]{}%
\providecommand \bibitemStop [0]{}%
\providecommand \bibitemNoStop [0]{.\EOS\space}%
\providecommand \EOS [0]{\spacefactor3000\relax}%
\providecommand \BibitemShut  [1]{\csname bibitem#1\endcsname}%
\let\auto@bib@innerbib\@empty
\bibitem [{\citenamefont {Coyne}\ and\ \citenamefont
  {Orr}(2004)}]{CoyneOrrSpeciationBook2004}%
  \BibitemOpen
  \bibfield  {author} {\bibinfo {author} {\bibfnamefont {J.~A.}\ \bibnamefont
  {Coyne}}\ and\ \bibinfo {author} {\bibfnamefont {H.~A.}\ \bibnamefont
  {Orr}},\ }\href@noop {} {\emph {\bibinfo {title} {Speciation}}}\ (\bibinfo
  {publisher} {Sinauer Associates, Inc.},\ \bibinfo {year} {2004})\BibitemShut
  {NoStop}%
\bibitem [{\citenamefont {Rosenzweig}(2001)}]{Rosenzweig2001}%
  \BibitemOpen
  \bibfield  {author} {\bibinfo {author} {\bibfnamefont {M.~L.}\ \bibnamefont
  {Rosenzweig}},\ }\href {\doibase 10.1073/pnas.101092798} {\bibfield
  {journal} {\bibinfo  {journal} {Proc Natl Acad Sci U S A}\ }\textbf {\bibinfo
  {volume} {98}},\ \bibinfo {pages} {5404} (\bibinfo {year}
  {2001})}\BibitemShut {NoStop}%
\bibitem [{\citenamefont {Darwin}(1859)}]{Darwin1859}%
  \BibitemOpen
  \bibfield  {author} {\bibinfo {author} {\bibfnamefont {C.~R.}\ \bibnamefont
  {Darwin}},\ }\href@noop {} {\emph {\bibinfo {title} {The Origin of
  Species}}}\ (\bibinfo  {publisher} {J. Murray, London},\ \bibinfo {year}
  {1859})\BibitemShut {NoStop}%
\bibitem [{\citenamefont {Dobzhansky}(1936)}]{Dobzhansky1936a}%
  \BibitemOpen
  \bibfield  {author} {\bibinfo {author} {\bibfnamefont {T.}~\bibnamefont
  {Dobzhansky}},\ }\href@noop {} {\bibfield  {journal} {\bibinfo  {journal}
  {Genetics}\ }\textbf {\bibinfo {volume} {21}},\ \bibinfo {pages} {113}
  (\bibinfo {year} {1936})}\BibitemShut {NoStop}%
\bibitem [{\citenamefont {Muller}(1942)}]{Muller1942}%
  \BibitemOpen
  \bibfield  {author} {\bibinfo {author} {\bibfnamefont {H.}~\bibnamefont
  {Muller}},\ }\href@noop {} {\bibfield  {journal} {\bibinfo  {journal} {Biol.
  Symp.}\ }\textbf {\bibinfo {volume} {6}},\ \bibinfo {pages} {71} (\bibinfo
  {year} {1942})}\BibitemShut {NoStop}%
\bibitem [{\citenamefont {Bateson}(1909)}]{Bateson1909}%
  \BibitemOpen
  \bibfield  {author} {\bibinfo {author} {\bibfnamefont {W.}~\bibnamefont
  {Bateson}},\ }\enquote {\bibinfo {title} {Darwin and modern science},}\ \
  (\bibinfo  {publisher} {Cambridge University Press},\ \bibinfo {year}
  {1909})\ pp.\ \bibinfo {pages} {85--101}\BibitemShut {NoStop}%
\bibitem [{\citenamefont {Mayr}(1963)}]{MayrBook1963}%
  \BibitemOpen
  \bibfield  {author} {\bibinfo {author} {\bibfnamefont {E.}~\bibnamefont
  {Mayr}},\ }\href@noop {} {\emph {\bibinfo {title} {Animal Species and
  Evolution}}}\ (\bibinfo  {publisher} {Harvard University Press, Cambridge,
  Mass.},\ \bibinfo {year} {1963})\BibitemShut {NoStop}%
\bibitem [{\citenamefont {Wu}\ and\ \citenamefont {Beckenbach}(1983)}]{Wu1983}%
  \BibitemOpen
  \bibfield  {author} {\bibinfo {author} {\bibfnamefont {C.~I.}\ \bibnamefont
  {Wu}}\ and\ \bibinfo {author} {\bibfnamefont {A.~T.}\ \bibnamefont
  {Beckenbach}},\ }\href@noop {} {\bibfield  {journal} {\bibinfo  {journal}
  {Genetics}\ }\textbf {\bibinfo {volume} {105}},\ \bibinfo {pages} {71}
  (\bibinfo {year} {1983})}\BibitemShut {NoStop}%
\bibitem [{\citenamefont {Vigneault}\ and\ \citenamefont
  {Zouros}(1986)}]{Vigneault1986}%
  \BibitemOpen
  \bibfield  {author} {\bibinfo {author} {\bibfnamefont {G.}~\bibnamefont
  {Vigneault}}\ and\ \bibinfo {author} {\bibfnamefont {E.}~\bibnamefont
  {Zouros}},\ }\href@noop {} {\bibfield  {journal} {\bibinfo  {journal}
  {Evolution}\ }\textbf {\bibinfo {volume} {40}},\ \bibinfo {pages} {1160}
  (\bibinfo {year} {1986})}\BibitemShut {NoStop}%
\bibitem [{\citenamefont {Orr}(1995)}]{Orr1995}%
  \BibitemOpen
  \bibfield  {author} {\bibinfo {author} {\bibfnamefont {H.~A.}\ \bibnamefont
  {Orr}},\ }\href@noop {} {\bibfield  {journal} {\bibinfo  {journal}
  {Genetics}\ }\textbf {\bibinfo {volume} {139}},\ \bibinfo {pages} {1805}
  (\bibinfo {year} {1995})}\BibitemShut {NoStop}%
\bibitem [{\citenamefont {Orr}\ and\ \citenamefont {Turelli}(2001)}]{Orr2001}%
  \BibitemOpen
  \bibfield  {author} {\bibinfo {author} {\bibfnamefont {H.~A.}\ \bibnamefont
  {Orr}}\ and\ \bibinfo {author} {\bibfnamefont {M.}~\bibnamefont {Turelli}},\
  }\href@noop {} {\bibfield  {journal} {\bibinfo  {journal} {Evolution}\
  }\textbf {\bibinfo {volume} {55}},\ \bibinfo {pages} {1085} (\bibinfo {year}
  {2001})}\BibitemShut {NoStop}%
\bibitem [{\citenamefont {Orr}\ and\ \citenamefont {Orr}(1996)}]{OrrOrr1996}%
  \BibitemOpen
  \bibfield  {author} {\bibinfo {author} {\bibfnamefont {H.}~\bibnamefont
  {Orr}}\ and\ \bibinfo {author} {\bibfnamefont {L.}~\bibnamefont {Orr}},\
  }\href@noop {} {\bibfield  {journal} {\bibinfo  {journal} {Evolution}\
  }\textbf {\bibinfo {volume} {50}},\ \bibinfo {pages} {1742} (\bibinfo {year}
  {1996})}\BibitemShut {NoStop}%
\bibitem [{\citenamefont {Gavrilets}(2003)}]{Gavrilets2003}%
  \BibitemOpen
  \bibfield  {author} {\bibinfo {author} {\bibfnamefont {S.}~\bibnamefont
  {Gavrilets}},\ }\href@noop {} {\bibfield  {journal} {\bibinfo  {journal}
  {Evolution}\ }\textbf {\bibinfo {volume} {57}},\ \bibinfo {pages} {2197}
  (\bibinfo {year} {2003})}\BibitemShut {NoStop}%
\bibitem [{\citenamefont {Fontana}(2002)}]{Fontana2002}%
  \BibitemOpen
  \bibfield  {author} {\bibinfo {author} {\bibfnamefont {W.}~\bibnamefont
  {Fontana}},\ }\href {\doibase 10.1002/bies.10190} {\bibfield  {journal}
  {\bibinfo  {journal} {Bioessays}\ }\textbf {\bibinfo {volume} {24}},\
  \bibinfo {pages} {1164} (\bibinfo {year} {2002})}\BibitemShut {NoStop}%
\bibitem [{\citenamefont {Force}\ \emph {et~al.}(1999)\citenamefont {Force},
  \citenamefont {Lynch}, \citenamefont {Pickett}, \citenamefont {Amores},
  \citenamefont {Yan},\ and\ \citenamefont {Postlethwait}}]{Force1999}%
  \BibitemOpen
  \bibfield  {author} {\bibinfo {author} {\bibfnamefont {A.}~\bibnamefont
  {Force}}, \bibinfo {author} {\bibfnamefont {M.}~\bibnamefont {Lynch}},
  \bibinfo {author} {\bibfnamefont {F.~B.}\ \bibnamefont {Pickett}}, \bibinfo
  {author} {\bibfnamefont {A.}~\bibnamefont {Amores}}, \bibinfo {author}
  {\bibfnamefont {Y.~L.}\ \bibnamefont {Yan}}, \ and\ \bibinfo {author}
  {\bibfnamefont {J.}~\bibnamefont {Postlethwait}},\ }\href@noop {} {\bibfield
  {journal} {\bibinfo  {journal} {Genetics}\ }\textbf {\bibinfo {volume}
  {151}},\ \bibinfo {pages} {1531} (\bibinfo {year} {1999})}\BibitemShut
  {NoStop}%
\bibitem [{\citenamefont {Berg}\ \emph {et~al.}(2004)\citenamefont {Berg},
  \citenamefont {Willmann},\ and\ \citenamefont {L\"{a}ssig}}]{Berg2004}%
  \BibitemOpen
  \bibfield  {author} {\bibinfo {author} {\bibfnamefont {J.}~\bibnamefont
  {Berg}}, \bibinfo {author} {\bibfnamefont {S.}~\bibnamefont {Willmann}}, \
  and\ \bibinfo {author} {\bibfnamefont {M.}~\bibnamefont {L\"{a}ssig}},\
  }\href {\doibase 10.1186/1471-2148-4-42} {\bibfield  {journal} {\bibinfo
  {journal} {BMC Evol Biol}\ }\textbf {\bibinfo {volume} {4}},\ \bibinfo
  {pages} {42} (\bibinfo {year} {2004})}\BibitemShut {NoStop}%
\bibitem [{\citenamefont {Khatri}\ \emph {et~al.}(2009)\citenamefont {Khatri},
  \citenamefont {McLeish},\ and\ \citenamefont {Sear}}]{Khatri2009}%
  \BibitemOpen
  \bibfield  {author} {\bibinfo {author} {\bibfnamefont {B.~S.}\ \bibnamefont
  {Khatri}}, \bibinfo {author} {\bibfnamefont {T.~C.~B.}\ \bibnamefont
  {McLeish}}, \ and\ \bibinfo {author} {\bibfnamefont {R.~P.}\ \bibnamefont
  {Sear}},\ }\href {\doibase 10.1073/pnas.0812260106} {\bibfield  {journal}
  {\bibinfo  {journal} {Proc Natl Acad Sci U S A}\ }\textbf {\bibinfo {volume}
  {106}},\ \bibinfo {pages} {9564} (\bibinfo {year} {2009})}\BibitemShut
  {NoStop}%
\bibitem [{\citenamefont {Iwasa}(1988)}]{Iwasa1988}%
  \BibitemOpen
  \bibfield  {author} {\bibinfo {author} {\bibfnamefont {Y.}~\bibnamefont
  {Iwasa}},\ }\href {\doibase 10.1016/S0022-5193(88)80243-1} {\bibfield
  {journal} {\bibinfo  {journal} {Journal of Theoretical Biology}\ }\textbf
  {\bibinfo {volume} {135}},\ \bibinfo {pages} {265 } (\bibinfo {year}
  {1988})}\BibitemShut {NoStop}%
\bibitem [{\citenamefont {Sella}\ and\ \citenamefont
  {Hirsh}(2005)}]{Sella2005}%
  \BibitemOpen
  \bibfield  {author} {\bibinfo {author} {\bibfnamefont {G.}~\bibnamefont
  {Sella}}\ and\ \bibinfo {author} {\bibfnamefont {A.~E.}\ \bibnamefont
  {Hirsh}},\ }\href {\doibase 10.1073/pnas.0501865102} {\bibfield  {journal}
  {\bibinfo  {journal} {Proc Natl Acad Sci U S A}\ }\textbf {\bibinfo {volume}
  {102}},\ \bibinfo {pages} {9541} (\bibinfo {year} {2005})}\BibitemShut
  {NoStop}%
\bibitem [{Note1()}]{Note1}%
  \BibitemOpen
  \bibinfo {note} {A more exact treatment using the Kramer's Moyal expansion
  from a generalised Master Equation gives the same Smoluchowski Equation, but
  for brevity is not shown.}\BibitemShut {Stop}%
\bibitem [{\citenamefont {Gardiner}(2009)}]{GardinerBook}%
  \BibitemOpen
  \bibfield  {author} {\bibinfo {author} {\bibfnamefont {C.}~\bibnamefont
  {Gardiner}},\ }\href@noop {} {\emph {\bibinfo {title} {Stochastic Methods: A
  Handbook for the Natural and Social Sciences}}}\ (\bibinfo  {publisher}
  {Springer},\ \bibinfo {year} {2009})\BibitemShut {NoStop}%
\bibitem [{\citenamefont {van Kampen}(1981)}]{vanKampen}%
  \BibitemOpen
  \bibfield  {author} {\bibinfo {author} {\bibfnamefont {N.}~\bibnamefont {van
  Kampen}},\ }\href@noop {} {\emph {\bibinfo {title} {Stochastic Processes in
  Physics and Chemistry}}}\ (\bibinfo  {publisher} {North-Holland},\ \bibinfo
  {year} {1981})\BibitemShut {NoStop}%
\bibitem [{\citenamefont {Bedford}\ and\ \citenamefont
  {Hartl}(2009)}]{Bedford2009}%
  \BibitemOpen
  \bibfield  {author} {\bibinfo {author} {\bibfnamefont {T.}~\bibnamefont
  {Bedford}}\ and\ \bibinfo {author} {\bibfnamefont {D.~L.}\ \bibnamefont
  {Hartl}},\ }\href {\doibase 10.1073/pnas.0812009106} {\bibfield  {journal}
  {\bibinfo  {journal} {Proc Natl Acad Sci U S A}\ }\textbf {\bibinfo {volume}
  {106}},\ \bibinfo {pages} {1133} (\bibinfo {year} {2009})}\BibitemShut
  {NoStop}%
\bibitem [{\citenamefont {von Hippel}\ and\ \citenamefont
  {Berg}(1986)}]{Hippel1986}%
  \BibitemOpen
  \bibfield  {author} {\bibinfo {author} {\bibfnamefont {P.~H.}\ \bibnamefont
  {von Hippel}}\ and\ \bibinfo {author} {\bibfnamefont {O.~G.}\ \bibnamefont
  {Berg}},\ }\href@noop {} {\bibfield  {journal} {\bibinfo  {journal} {Proc
  Natl Acad Sci U S A}\ }\textbf {\bibinfo {volume} {83}},\ \bibinfo {pages}
  {1608} (\bibinfo {year} {1986})}\BibitemShut {NoStop}%
\bibitem [{\citenamefont {Gerland}\ \emph {et~al.}(2002)\citenamefont
  {Gerland}, \citenamefont {Moroz},\ and\ \citenamefont {Hwa}}]{Gerland2002a}%
  \BibitemOpen
  \bibfield  {author} {\bibinfo {author} {\bibfnamefont {U.}~\bibnamefont
  {Gerland}}, \bibinfo {author} {\bibfnamefont {J.~D.}\ \bibnamefont {Moroz}},
  \ and\ \bibinfo {author} {\bibfnamefont {T.}~\bibnamefont {Hwa}},\ }\href
  {\doibase 10.1073/pnas.192693599} {\bibfield  {journal} {\bibinfo  {journal}
  {Proc Natl Acad Sci U S A}\ }\textbf {\bibinfo {volume} {99}},\ \bibinfo
  {pages} {12015} (\bibinfo {year} {2002})}\BibitemShut {NoStop}%
\bibitem [{Note2()}]{Note2}%
  \BibitemOpen
  \bibinfo {note} {Note that for larger populations discrete dynamics will come
  into play as on each lineage the populations distributions become confined to
  within a single hamming distance of the zero hamming distance state. This
  state has zero degeneracy and so the two lineages can only diverge via
  substitutions to larger, lower fitness, hamming distances at an exponentially
  reduced rate compared to neutrality.}\BibitemShut {Stop}%
\end{thebibliography}
%

\end{document}